\documentclass{article}
\usepackage{graphicx}
\usepackage{mathrsfs}
\usepackage{float}
\usepackage{subfig}
\usepackage{caption}
\usepackage{bbm}
\usepackage{amsmath}
\usepackage{algorithm2e}
\RestyleAlgo{ruled}
\DontPrintSemicolon

\newtheorem{definition}{Definition}[section]

\PassOptionsToPackage{numbers}{natbib}

\usepackage[preprint]{neurips_2024}

\usepackage[utf8]{inputenc} 
\usepackage[T1]{fontenc}    
\usepackage{hyperref}       
\usepackage{url}            
\usepackage{booktabs}       
\usepackage{amsfonts}       
\usepackage{nicefrac}       
\usepackage{microtype}      
\usepackage{xcolor}         

\title{Temporal fingerprints: \\Identity matching across fully encrypted domains}

\author{%
  Shahar Somin\\
  MIT Media Lab\\
  Cambridge, MA 02139 \\
  \texttt{shaharso@mit.edu} \\
  \And
  Keeley Erhardt \\
MIT Media Lab\\
   Cambridge, MA 02139 \\
  \texttt{keeley@mit.edu} \\
  \And
  Alex Pentland \\
  MIT Media Lab\\
   Cambridge, MA 02139 \\
  \texttt{pentland@mit.edu} \\
}

\begin{document}

\maketitle

\begin{abstract}
  Technological advancements have significantly transformed communication patterns, introducing a diverse array of online platforms, thereby prompting individuals to use multiple profiles for different domains and objectives.
Enhancing the understanding of cross domain identity matching capabilities is essential, not only for practical applications such as commercial strategies and cybersecurity measures, but also for theoretical insights into the privacy implications of data disclosure. 
In this study, we demonstrate that individual temporal data, in the form of inter-event times distribution, constitutes an individual temporal fingerprint, allowing for matching profiles across different domains back to their associated real-world entity. 
We evaluate our methodology on encrypted digital trading platforms within the Ethereum Blockchain and present impressing results in matching identities across these privacy-preserving domains, while outperforming previously suggested models.
Our findings indicate that simply knowing \textit{when} an individual is active, even if information about \textit{who} they talk to and \textit{what} they discuss is lacking, poses risks to users' privacy, highlighting the inherent challenges in preserving privacy in today's digital landscape. 

\end{abstract}

\section{Introduction}

In an era characterized by digital ubiquity, the nature of human interaction has undergone a profound transformation. 
Historically, our interactions, whether social or commercial, relied on physical encounters, bound to a singular persona --- our physical identity.
However, with the emergence of online platforms and digital technologies, the vast majority of our activities occur online, allowing us to conduct them from the comfort of our own homes, while facilitating the use of different profiles for diverse objectives and platforms.
Given these transformations, identifying different profiles associated with the same individual across distinct domains becomes pivotal.
This capability is essential for numerous practical aspects, spanning both commercial interests like user-oriented recommendation systems \cite{yan2015unified} and cyber-security considerations, encompassing identity verification, fraud detection \cite{kasa2019improving}, impersonations and cyber attacks \cite{li2011identity,viswanath2015strength,kumar2017army}.
Beyond these practical applications, delving into the cross domain identity matching problem offers significant insights into the ramifications of data disclosure on individual privacy.

Extensive research efforts have demonstrated that both content and personal attributes are inherently sensitive data in terms of individual privacy \cite{GDPR}. 
Indeed, techniques leveraging user-generated content \cite{wang2018you,tai2019adversarial} or personal profile characteristics such as age, gender or usernames  \cite{perito2011unique,liu2013s,mu2016user,labitzke2011your} have presented notable efficacy in detecting profiles corresponding to the same individual across different domains.
Furthermore, even methodologies applied to anonymized domains, have exhibited remarkable cross-domain identity matching capabilities, highlighting the privacy implications of granting access to the network of interactions \cite{heimann2018regal,singh2008global,derr2021deep,nassar2018low,crectu2022interaction}
 and to individual metadata, such as profile trajectories \cite{riederer2016linking}.

This raises an intriguing question --- what is the level of privacy assurances achievable in anonymous domains sharing only individual temporal data, while restricting the access to both the content and the network of interactions?
Indeed, the temporal traces we leave behind, ranging from the timing of phone calls, digital trading or our activity over social platforms are ubiquitous.
Over the years, the temporal analysis of such traces has proven useful in a variety of applications.
These include charactering group dynamics \cite{raghavan2013hidden} and predicting events such as terror attacks \cite{clauset2013estimating} and epidemic outbursts \cite{scarpino2019predictability}.
Furthermore, techniques based on temporal data have demonstrated the ability to differentiate individuals 
using the dynamics of their keystroke timings \cite{banerjee2012biometric}, detect network intrusions based on user activity timing \cite{scott2001detecting}, and enabled identifying devices by analysing packet arrival time intervals from wireless apparatus \cite{radhakrishnan2014gtid}.

In this study, we present an unsupervised model for cross-domain identity matching which relies merely on temporal data at the individual-level.
The model defines an affinity function between profiles, based on the time-gaps between a profile's consecutive activities.
We demonstrate that the distribution of these \textit{inter-event} times constitutes an individual \textit{temporal fingerprint}.
Namely, we show that despite individuals utilizing different profiles in distinct domains, and often also acting in non-overlapping times, their activities exhibit a shared underlying pattern, encapsulated in their inter-event time distribution.

We demonstrate our methodology on encrypted digital trading platforms from the Ethereum Blockchain \cite{somin2022remaining, somin2022beyond, somin2020network,buterin2014next}, constituting of over 250k daily traders.
The identity matching problem across these domains establishes a notably low baseline, as merely 3 out of 1000 randomly selected profile pairs actually correspond to the same individual.
Nevertheless, our methodology was able to achieve an average AUC of $0.78$ and precision of $96\%$ for the top-100 predictions. 
We further demonstrate that even if structural data in the form of the interactions network is available for this task, it establishes significantly inferior performance compared to our temporal model. 
Finally, we show that our methodology presents high stability along time as well as significant robustness to noise.
Our results provide evidence that users remain identifiable even in encrypted systems, where on top of content restrictions, the underlying network of interactions is also lacking. 
These findings underscore the sensitivity of individual temporal data with regards to privacy considerations and highlight the inherent challenges in preserving privacy in the digital age.

\section{Preliminaries}\label{sec:methods}
In this section, we introduce the essential terminology, notations and definitions of our proposed approach.

\subsection{Problem definition}
In this study we aim at identifying individuals across different encrypted domains. 
Specifically, we aim at learning an identity matching function:
\begin{definition}\textbf{cross-domain identity matching function}
Given $D_1,...,D_n$ different domains, the goal of the cross-domain identity matching problem is learning a function:
\[
p: \bigcup_{i,j \in [n]}D_i \times D_j \rightarrow [0,1]
\]
such that  $p(u_{d_1},v_{d_2})$ represents the probability that the profiles $u_{d_1}\in D_1$ and $v_{d_2}\in D_2$ are associated with the same real-world individual ($u_{d_1} = v_{d_2}$).
\end{definition}

\subsection{The proposed model}
At the absence of content and network information we propose exploiting individual temporal data for linking profiles back to the same individual across different domains.
Specifically, we analyse the time difference between any two consecutive activities of each profile. Formally:
\begin{definition}
    Given a day $\tau$ and a profile $u_d$ in domain $D$ we denote 
    the sequence of their activity times  $A^{u_d}_{\tau}\subset \left[\tau,\tau+1\right]$ by:
    \begin{equation}
        A^{u_d}_{\tau} = (t_0^{u_d},...,t_m^{u_d})
    \end{equation}
    An inter-event time period is defined as the time difference between two consecutive activities of $u_d$:
    \begin{equation}
        \Delta t_i^{u_d} = t_{i}^{u_d} - t_{i-1}^{u_d}
    \end{equation}
    The inter-event time sequence is defined by:
    \begin{equation}
        S^{u_d}_{\tau}=(\Delta t_1^{u_d},...\Delta t_m^{u_d})
    \end{equation}
    The cumulative distribution function of the inter-event sequence is defined as:
    \begin{equation}
        Q^{u_d}_{\tau}(\Delta t) = \frac{|\delta \in S^{u_d}_{\tau}: \delta \leq \Delta t|}{m}
    \end{equation} 
\end{definition}

Our suggested identity matching function $p^{ks}$ is based on the similarity between the established inter-event time distributions of any two profiles.
In order to estimate the similarity of two distributions we perform the two-sample Kolmogorov-Smirnov test for goodness-of-fit. 

\begin{definition}
Let $u_{d_1}\in D_1^{\tau}$ and $v_{d_2} \in D_2^{\tau}$, and their corresponding inter-event time distributions $Q^{u_{d_1}}_{\tau}$ and $Q^{v_{d_2}}_{\tau}$. 
The KS-statistic is defined as the maximal difference between their distributions:
\begin{equation}
    KS_{\tau}(u_{d_1},v_{d_2})= \sup_{\Delta t} |Q^{u_{d_1}}_{\tau}(\Delta t)-Q^{v_{d_2}}_{\tau}(\Delta t)|
\end{equation}
The null hypothesis of the two-sided Kolmogorov-Smirnov goodness-of-fit test states that both samples originate from the same distribution.
The null hypothesis is rejected with significance level $\alpha$ if:
\begin{equation}
    KS_{\tau} (u_{d_1},v_{d_2})> \sqrt{-\ln{(\frac{\alpha}{2})}\cdot \frac{1+\frac{m}{k}}{2m}}
\end{equation}
where $m=|A_{\tau}^{u_{d_1}}|$ and $k=|A_{\tau}^{v_{d_2	}}|$.
We define $\mathbbm {1}_{gof}$ to indicate when the null hypothesis can not be rejected:
\begin{equation}
     \mathbbm {1}_{gof}(u_{d_1},v_{d_2}):=
  \begin{cases}1~&{\text{ if }}~KS_{\tau}(u_{d_1},v_{d_2})\leq \sqrt{-\ln{(\frac{\alpha}{2})}\cdot \frac{1+\frac{m}{k}}{2m}}~,\\0~&{\text{ else }}
  \end{cases}
\end{equation}
The corresponding identity matching function $p^{ks}$ first sorts profile pairs according to $\mathbbm {1}_{gof}$ and performs a secondary sort by the actual KS distances.  
\end{definition}

\subsection{Temporal graph neural network model}
\begin{figure}[H]
		\centering
		{\includegraphics[height=3.cm]{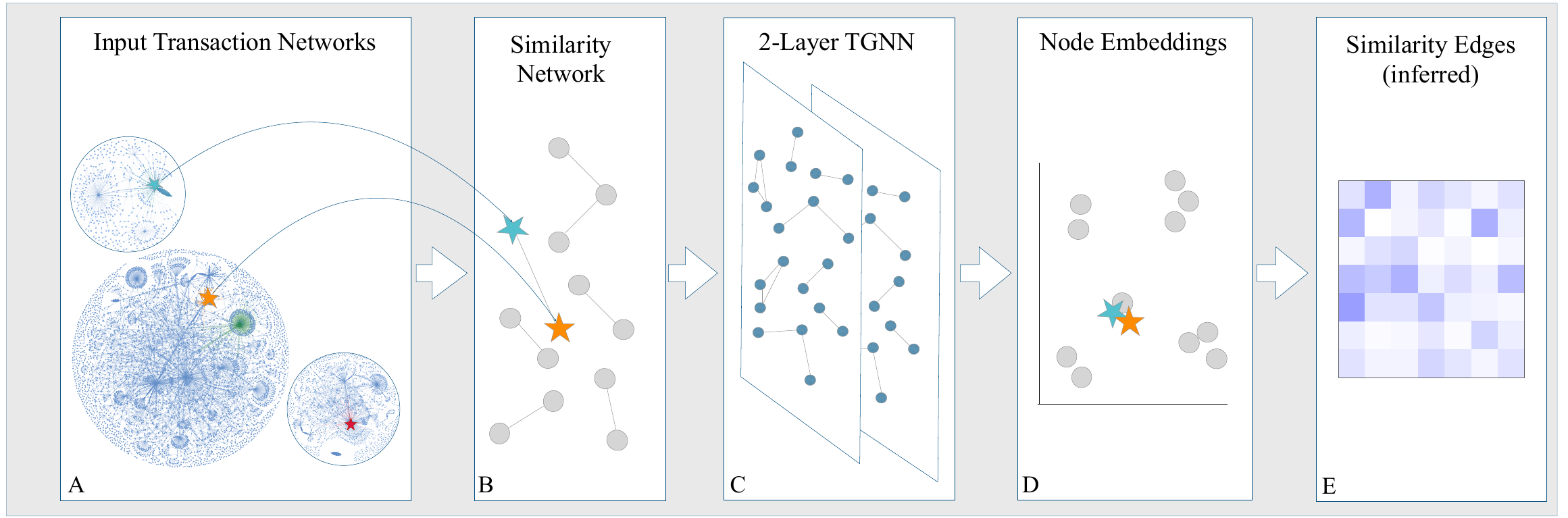} }
		\caption{Temporal Graph Neural Network (TGNN) flow. 
		Panel A presents the initial daily transaction networks, encorporating timings of individual node activities. 
		Panel B depicts the similarity networks induced from the daily inter-event distributions similarities. 
		A 2-layer TGNN is trained on positive and negative edges (panel C) to produce a latent node embedding (panel D), which is used to estimate the similarity between two nodes (panel E).
	}
		\label{fig:fig5_gnn}
	\end{figure}

To further enhance this vanilla version of inter-event-based synchronization, we propose employing a temporal graph neural network (TGNN) model on top of cross-domain \textit{similarity networks}  $G_{ks}^{\tau} = (V^{\tau},\,E^{\tau})$ where edges link profiles across different domains,
and edge weight is the Kolmogorov-Smirnov distance between the inter-event time distributions of any pair of profiles. 
We employ a supervised learning approach\footnote{The proposed TGNN setting utilizes labels inferred from the KS distance and does not rely on genuine identity labels, as such it is applicable to the unsupervised approach we are testing.} to train a TGNN, using edge labels inferred from the KS distance metric: 
 \begin{enumerate}
    \item \textbf{Positive edges:} two profiles $u_{d_1}\in D_1^{\tau}$ and $v_{d_2}\in D_2^{\tau}$ are linked by a positive edge if $KS_{\tau}(u_{d_1},v_{d_2})\leq \rho^p$, where $\rho^p$ is a predefined positive threshold.
    \item \textbf{Negative edges:} two profiles $u_{d_1} \in D_1^{\tau}$ and $v_{d_2}\in D_2^{\tau}$ are linked by a negative edge if $KS_{\tau}(u_{d_1},v_{d_2})\geq \rho^n$, where $\rho^n$ is a predefined negative threshold.
\end{enumerate}
The TGNN learns a latent embedding for all profiles, which is utilized subsequently for a cross-domain edge detection task. 
The 2-layer TGNN employed on the daily similarity networks is illustrated in Fig. \ref{fig:fig5_gnn} (elaborated TGNN architecture details can be found in the SI).

\subsection{Experiments}
We examine our methodology on transactional data from the Ethereum Blockchain.
This fully encrypted ecosystem enables the trading of tens of thousands of different crypto-tokens, using a single Ethereum wallet.
Broadly, a crypto wallet is a digital tool that securely stores and manages the user's cryptocurrency holdings, allowing them to send, receive, and monitor their digital assets on blockchain networks. 
The address of a crypto wallet serves as a unique identifier, similarly to an account number in traditional financial systems. 
Since a single Ethereum wallet can be employed for trading of all Ethereum-based crypto-tokens, it can be used as the trader's identifier across the different crypto-domains, for validation purposes.
Consequently, we split this financial ecosystem into multiple crypto-domains, such that each crypto-domain $D^i_{\tau}$ encompasses all the trading activity related to the respective crypto-token $c_i$ and day $\tau$:
\begin{equation}
        D^i_{\tau} = \{u : u \text{ bought or sold } c_i \text{ on day }\tau \}
    \end{equation}


We compare the proposed identity function to a naive rule measuring the activity overlap between two profiles. Formally:
\begin{definition}
Given  $u_{d_1}\in D^1_{\tau}$ and $v_{d_2} \in D^2_{\tau}$, and their corresponding activity times $ A^{u_{d_1}}_{\tau} = (t_0^{u_{d_1}},...,t_m^{u_{d_1}})
 $ and $ A^{u_{d_2}}_{\tau} = (t_0^{u_{d_2}},...,t_m^{u_{d_2}})
 $, we define the identity matching function based on their activity overlap by:
\begin{equation}\label{eq:activity_overlap}
  p^{ao}_{\tau}(u_{d_1},v_{d_2}) = min \{\frac{|A^{u_{d_1}}_{\tau}\cap A^{u_{d_2}}_{\tau}|}{|A^{u_{d_1}}_{\tau}|}, \frac{|A^{u_{d_1}}_{\tau}\cap A^{u_{d_2}}_{\tau}|}{|A^{u_{d_2}}_{\tau}|}\}
\end{equation}
\end{definition}

We further compare the temporal identity matching functions to a structural dependent one. We utilize the REGAL \cite{heimann2018regal} 
network alignment algorithm, which first establishes a network embedding based on the transactions network 
and then employs a nearest neighbor 
search for finding node alignments. 
The corresponding identity matching function $p^{regal}_{\tau}$ as:
\begin{equation}
p^{regal}_{\tau}(u_{d_1},v_{d_2})\propto euc \big(emb_{\tau}(u_{d_1}),emb_{\tau}(v_{d_2})\big)^{-1}
\end{equation}
where $emb_{\tau}(u)$ is the REGAL-based embedding of node $u$ based on its transaction network on day ${\tau}$, and $euc(\cdot,\cdot)$ is the Euclidean distance between node embeddings. 

Finally, we examine the effect of temporality on the identity-matching problem. 
To this end we employ the TGNN model to the temporal similarity networks, training on positive edges defined by $\rho^p=0.001$ and negative edges defined by $\rho^n=0.98$. 
The model is evaluated on a validation network containing all edges $(u_{d_1},v_{d_2})$ having $0.001<KS(u_{d_1},v_{d_2})\leq 0.02616$ (standing for the next top KS $1000$ edges). 
The corresponding identity-matching function $p^{tgnn}$ as follows: \begin{equation}
p^{tgnn}(u_{d_1},v_{d_2}) = \mathfrak{f}\big(emb_{tgnn}(u_{d_1}),emb_{tgnn}(v_{d_2})\big)
\end{equation}  
where $\mathfrak{f}$ is the learnt classifier.

\subsection{Evaluation}
For a given day $\tau$, we denote the set of all correct profile pairs by:
\begin{equation}
\mathcal{P}_{\tau} = \{(u_{d_i},v_{d_j})\in \bigcup_{i,j \in [n]}D^i_{\tau} \times D^j_{\tau} \,\,|\,\,u_{d_i}=v_{d_j} \}
\end{equation}
We evaluate the performance of our suggested model by employing several standard methods; \textit{Area Under the ROC Curve (AUC)} and \textit{Precision}.
We also measure the \textit{Precision@k}, prevalent in identity-matching evaluations \cite{zhou2018deeplink,chu2019cross}, which evaluates a \textit{softer} matching requirement.
Specifically, we examine the top-$k$ candidate matches for each profile in all possible domains.
We denote the number of correct matches for $u$ by $c_{u}^{k} $ and: 
\begin{equation}
\textit{Precision@k} =\frac{\sum_u c_{u}^{k}}{l}
\end{equation}
where $l=|\mathcal{P}_{\tau} |$ is the number of correct profile pairs. 
We report the average performance in each of the above mentioned metrics, over $14$ days of tests, alongside with the established standard error.

\section{Results}

\begin{figure}[h]
		\centering
		{\includegraphics[height=4.68cm]{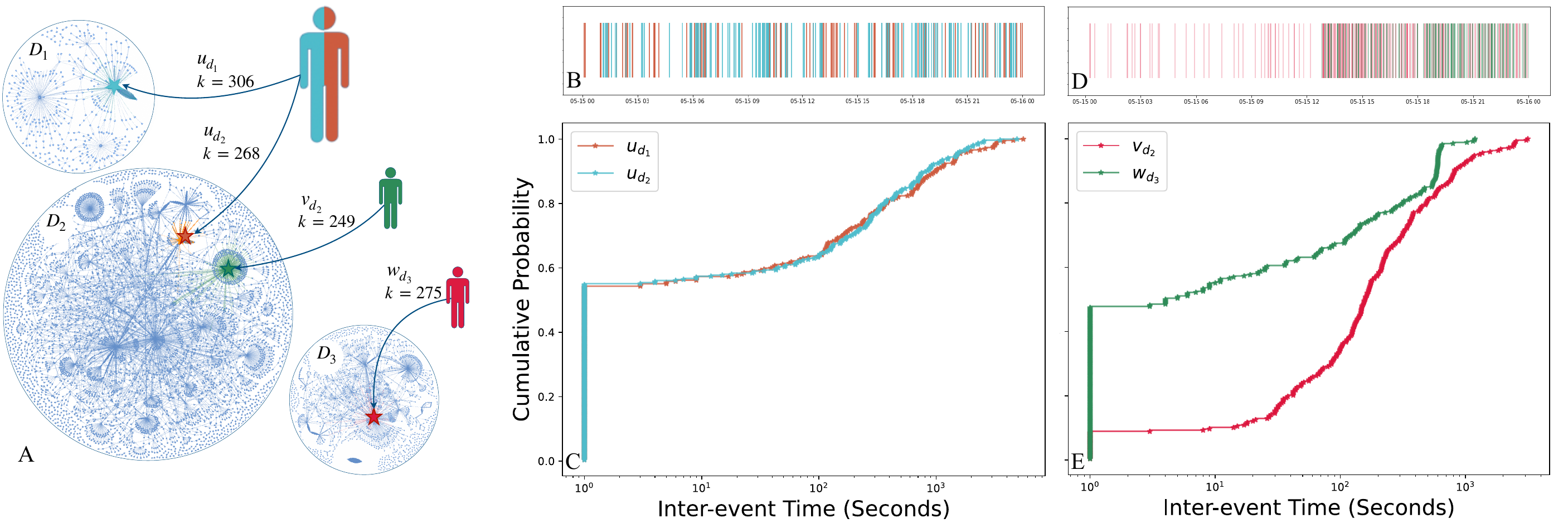} }
		\caption{Temporal synchronization of four examined profiles.
        Panel A illustrates the daily networks of three domains $D^1_{\tau}$, $D^2_{\tau}$ and $D^3_{\tau}$, each corresponding to the trading of a different crypto-token. 
        Profiles $u_{d_1}$ (degree 306) and $u_{d_2}$ (degree 268) correspond to the same individual (illustrated by orange and cyan markers).
        Profiles $v_{d_2}$ (degree 249) and $w_{d_3}$ (degree 275) pertain to different individuals (illustrated by red and green markers).
        Panel B presents the activity times of $u_{d_1}$ and
        $u_{d_2}$, reaching an activity overlap of $37\%$.
        Panel D presents the activity times of $v_{d_2}$ and $w_{d_3}$, reaching an activity overlap of $42\%$. 
        Panel C depicts the cumulative inter-event times distributions of $u_{d_1}$ and
        $u_{d_2}$, exemplifying similar distributions ($KS_{\tau}(u_{d_1},u_{d_2})=0.031$ with a p-value of $0.99$).
        Panel D depicts the cumulative inter-event times distributions of $v_{d_2}$ and $w_{d_3}$, exemplifying significantly different distributions ($KS_{\tau}(v_{d_2},w_{d_3})=0.47$ with a p-value of $5e-27$).}
		\label{fig:fig1_silences}
\end{figure}

We postulate that different profiles pertaining to the same individual are bound to exhibit synchronization in their activity dynamics, despite acting on different domains. 
Fig \ref{fig:fig1_silences} suggests that the inter-event time distribution holds greater promise for identity matching across different domains, compared to other temporal or structural node characteristics. 
We demonstrate this by the following example, observing two pairs of profiles:
\begin{enumerate}
    \item \label{def:positive_pair}A positive pair: $u_{d_1}$ and $u_{d_2}$, correspond to the same individual $u$ in domains $D^1_{\tau}$ and $D^2_{\tau}$ respectively, illustrated by orange and cyan markers in panel A, Fig. \ref{fig:fig1_silences}.
    \item A negative pair: $v_{d_2}$ and $w_{d_3}$, corresponding to different individuals in two different domains $v\in D^2_{\tau},\,\,w\in D^3_{\tau}$, illustrated by red and green markers in panel A, Fig. \ref{fig:fig1_silences}. 
\end{enumerate}

Both pairs present similar degrees (illustrated in panel A, Fig. \ref{fig:fig1_silences}) and resembling percentage of activity times overlap: $37\%$ of the time for the positive pair and $42\%$ for the negative pair, illustrated in panels B and D, Fig. \ref{fig:fig1_silences}, respectively. 
Nevertheless, the negative pair presents significantly different inter-event times distributions, establishing a Kolmogorov-Smirnov distance of $KS_{\tau}(v_{d_2},w_{d_3})=0.47$ with a p-value of $5e^{-27}$ (panel E, Fig. \ref{fig:fig1_silences}), while the positive pair presents a high similarity of the inter-event distributions, with $KS_{\tau}(u_{d_1},u_{d_2})=0.031$ and a p-value of $0.99$ (illustrated in panel C, Fig. \ref{fig:fig1_silences}).
Interestingly, this illustrates that different profiles associated with a single individual might manifest notable levels of synchronization despite presenting only a mild overlap in their activity times.

\begin{figure}[h]
		\centering
		{\includegraphics[height=5.4cm]{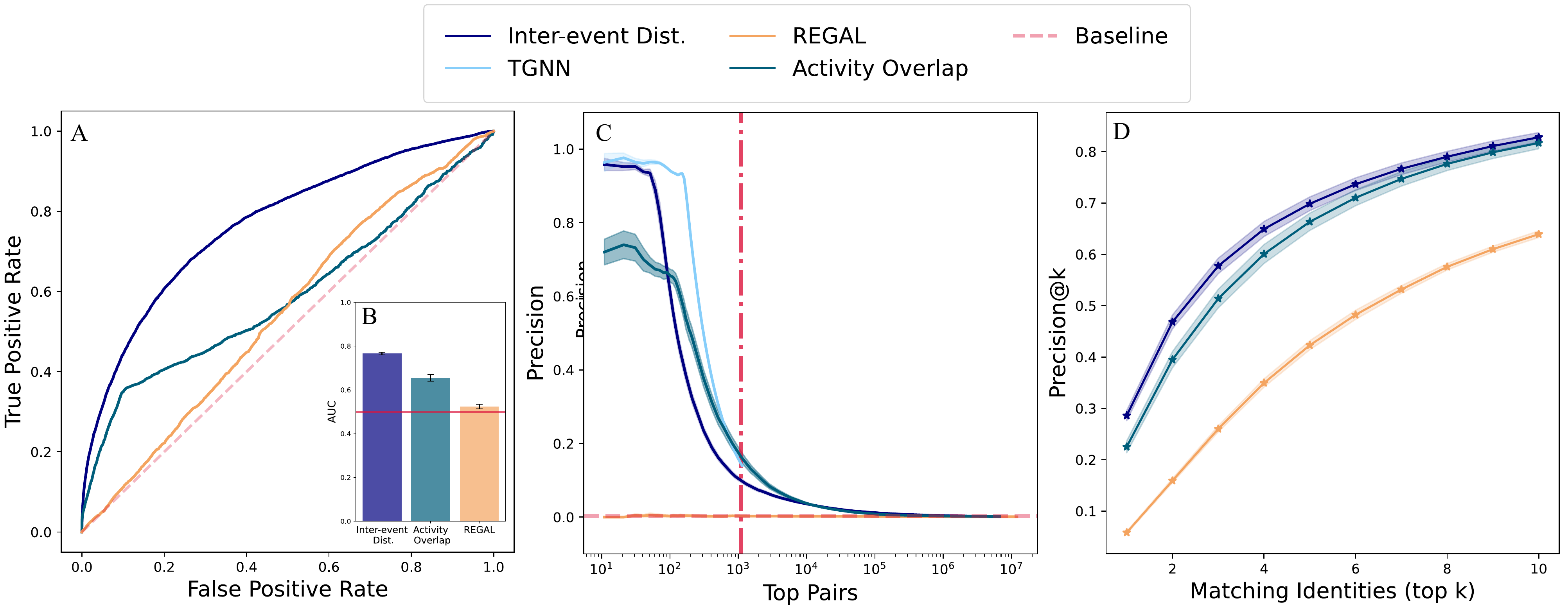} }
		\caption{Performance evaluation of the inter-event similarity method for identity matching and comparison to other temporal and structural baselines.
        Panel A depicts the ROC-curve established for matching identities on an arbitrary day of data for both $p^{ks}$ and $p^{ao}$ and panel B presents the averaged AUC over $14$ daily tests, with error bars standing for standard error.
        Panel C presents the average precision of $p^{ks}$ and $p^{ao}$ as a function of the examined number of pair candidates (with $\pm 1$ standard error in light background, correspondingly). 
        Panel D depicts the comparison of the precision@k metric across the examined identity-matching functions.
        The inter-event similarity method presents higher performance than baseline methods across all examined metrics. 
        The TGNN-based method manifests an evident enhancement to the top-1000 precision.
	}
		\label{fig:fig2_silences_perf}
	\end{figure}

We wish to evaluate the performance of the inter-event distribution-based method and compare it to other methods exploiting structural and temporal node characteristics ($p^{ao}$, $p^{regal}$ and $p^{tgnn}$). 
 Fig. \ref{fig:fig2_silences_perf} presents the performance analysis of these four methods. 
Specifically, we consider $14$ days of activity over the examined crypto-domains $D^1_{\tau},...,D^n_{\tau}$ and evaluate the performance on each day separately. 
Panel A. in Fig. \ref{fig:fig2_silences_perf} presents the ROC-curve achieved by $p^{ks}_{\tau}$ on an arbitrary day of data, manifesting a significantly higher AUC compared to the $p^{ao}_{\tau}$ and $p^{regal}_{\tau}$ baselines. 
In general, as seen in panel B, Fig. \ref{fig:fig2_silences_perf}, the daily AUC achieved by $p^{ks}$ presents higher values compared to the baselines $p^{ao}$ and $p^{regal}$, averaged on the examined $14$ days of data. 
Panel C, Fig. \ref{fig:fig2_silences_perf} presents the precision established for each threshold of top-ranked profile pairs $(u_{d_1},v_{d_2})$. 
Notably, up to the top $200$ pairs $p_{ks}$ significantly outperforms $p_{ao}$, reaching almost error-less precision on average. 
Panel D, Fig. \ref{fig:fig2_silences_perf} depicts the \textit{precision@k} of $p^{ks},\,p^{regal}$ and $p^{ao}$ for different values of $k$. 
It demonstrates that the inter-event model is able to identify $83\%$ of the correct pairs after examining at most $10$ candidates per profile, thereby outperforming the baseline models. 

We further wish to verify whether we can enhance the identity matching performance by addressing both the dynamics of the inter-event distributions and structural patterns from the original transaction networks. 
To this end we employ a TGNN over all $14$ daily snapshots and define structural and temporal node features (consider SI for elaborated view of the TGNN architecture). 
Light blue curve in panel D, Fig. \ref{fig:fig2_silences_perf} presents the precision the TGNN evaluated on the top-$1000$ KS similarity edges (consider section \ref{sec:discussion} for a discussion of the runtime complexity of the TGNN).
The evident performance boost to the vanilla KS model underscores the evolving role of each profile within the overall network.




These findings are highly promising, particularly when considering the exceptionally low baseline for the identity matching problem, which stands on $0.3\%$, signifying that only $3$ out of $1000$ random draws will result in a true match (depicted by the horizontal red line in panel C).
We conclude that although the inter-event times distribution is highly entangled with the actual activity times of an individual, the latter exhibits worse performance, thus indicating its inadequacy in providing an accurate individual fingerprint.
Additionally, we observe that the structure-based approach demonstrates limited efficacy in linking profiles across different domains, highlighting the greater significance of temporality in this context (consider section \ref{sec:discussion} for a thorough discussion).

We next wish to evaluate how noise induced into the system affects 
the ability to identify matching profiles across different domains
We define the \textit{noisy} setting as follows: 
\begin{definition}
Given a day $\tau$ and a profile $u_{d}$ in domain $D_{\tau}$, Gaussian noise $\mathcal{N}(\mu,\sigma^2)$ is added to each profile's activity times.
The corresponding noisy activity times are denoted by:
\begin{equation}
    \widehat{A^{u_d}_{\tau}} = A^{u_d}_{\tau}  + \mathcal{N}(\mu,\sigma^2)
\end{equation}
and the inter-event time distributions are denoted by $\widehat{Q^{u_d}_{\tau}}$.
\end{definition}

\begin{figure}[h]
		\centering
		{\includegraphics[height=5cm]{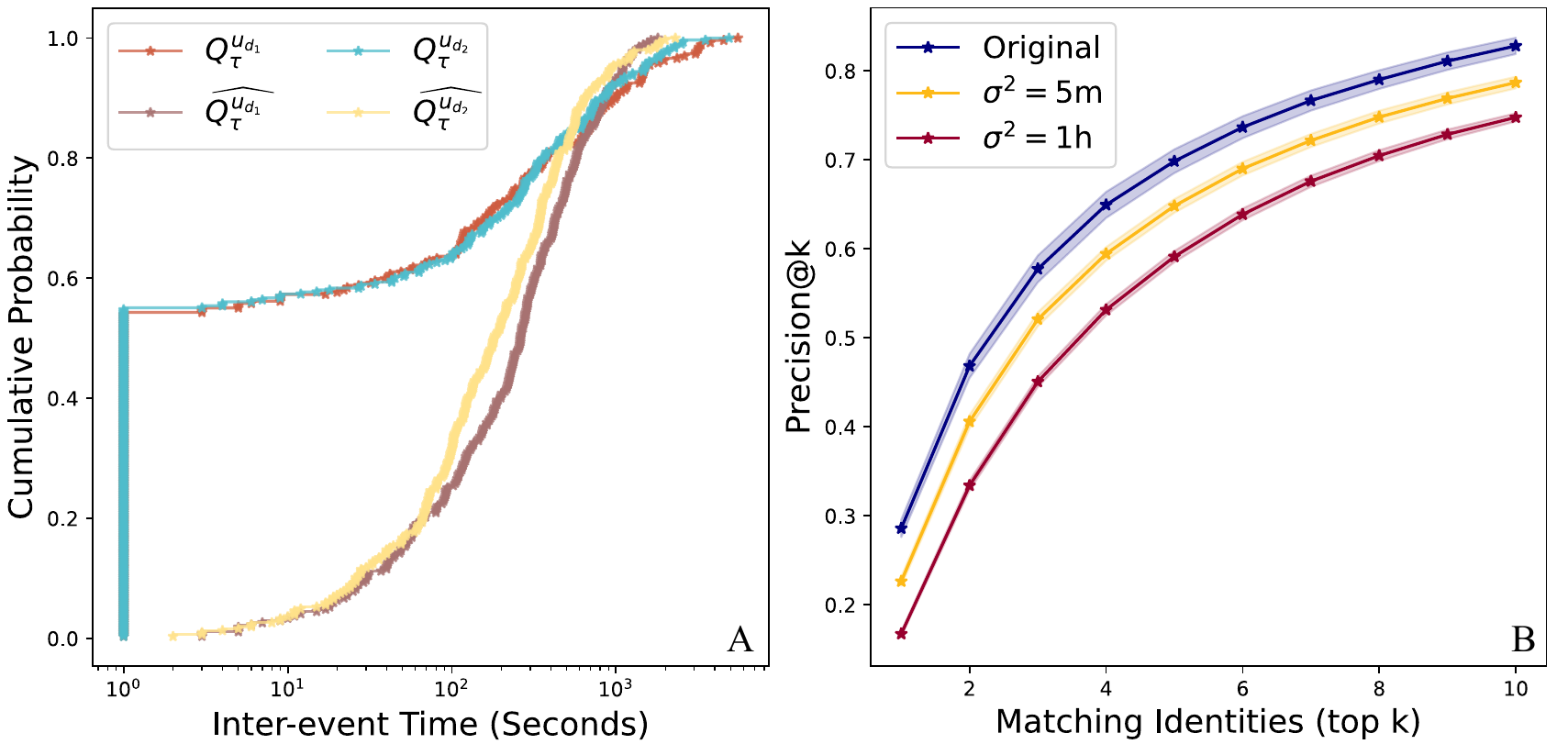} }
		\caption{Noise injection effect on profile similarity.
		Panel A depicts the cumulative inter-event time distributions of two examined profiles after injecting $\mathcal{N}(0,1h)$ of noise, compared to their original noiseless distributions. 
		Notably, the profiles remain similar despite the injection of noise.
Panel B depicts the precision@k metric 
as a function of the number of examined ranked matches (k) and its dependence on the levels of noise, illustrating that $78\%$ and $ 74\% $  (on average, $\pm 1$ standard error in light background) of the correct profile pairs are identifiable within merely $10$ ranks after the injection of $\mathcal{N}(0,5m)$ and $\mathcal{N}(0,1h)$ of noise, correspondingly.
	}
		\label{fig:fig4_robustness_noise}
	\end{figure}

We analyze the noisy setting with different amounts of injected noise. We fix $\mu=0$ and consider $\sigma^2=5$ and $60$ minutes. 
We start by revisiting the two profiles we have introduced in example \ref{def:positive_pair}. 
Panel A in Fig. \ref{fig:fig4_robustness_noise} depicts their cumulative inter-event time distributions after injecting $\mathcal{N}(0,5)$ of noise, compared to their original noiseless distributions. 
While for each profile, its pre- and post noise distributions change significantly, we note that the similarity between the two different profiles remains high even after the noise injection.
Consequently, we wish to estimate the noise injection impact on the performance of our identity matching function.
Panel B depicts the precision@k metric for each of the different noisy settings. 
Interestingly, it illustrates $80\% $ of the correct profile pairs are identifiable within merely $10$ ranks after the injection of $\mathcal{N}(0,5m)$ of noise.
This ascertains that even at relatively low costs, namely checking $10$ candidates per user, the suggested methodology is able to detect its associated profiles despite the injection of significant noise.

\section{Related Work}\label{sec:related_work}
\paragraph{Inter-event time distributions}
The distribution of inter-event times has been a subject of scientific research for the last two decades.
During these endeavours it was demonstrated that human temporal activity is characterized by bursts of activity and long pauses thereafter. 
This bursty nature of human activity is mathematically represented by a heavy tailed distribution of the inter-event times. 
These heavy-tailed inter-event time distributions were encountered in many and diverse types of activities, including individual mobility patterns, e-mail communications, instant messaging, web browsing and mobile phone calls \cite{barabasi2005origin,gonzalez2008understanding, karsai2011small,oliveira2005darwin, zhou2008role, dezso2006dynamics,candia2008uncovering}. 


\paragraph{Cross-domain identity matching}
Previous research efforts related to cross-domain identity matching problems, or by their other names as network alignment or user identity linkage, most frequently aimed to solve this problem utilizing supervised learning models.
These refer to the problem as a binary classification problem, where positive samples are $\mathcal{P} = \{(u_{d_i},v_{d_j})_{i,j\in [n]}
\,\, | \,\,u=v\}$  and negative samples are  $\mathcal{N} = \{(u_{d_i},v_{d_j})_{i,j\in [n]}
\,\, | \,\,u\neq v\}$. 
These samples are divided into training and test sets, and the supervised model aim to learn an identity matching function $p$ on the training test, whose performance is evaluated on the test set. 
An extensive research effort has been directed towards supervised learning of cross domain linkage, including probabilistic methods \cite{perito2011unique, zhang2014online} and embedding-based methods \cite{mu2016user,fey2020deep}.
While most supervised models heavily rely on user attributes \cite{zafarani2013connecting,malhotra2012studying,goga2013large}
several researches have shown great prominence using merely the network of interactions
 \cite{man2016predict,zhou2018deeplink,chu2019cross}.
 
 On the other hand, unsupervised models do not require labels samples for learning. 
 They often are based on some similarity measure computed on each candidate pair of user profiles. 
 For instance, cite{riederer2016linking} utilized user trajectories and \cite{labitzke2011your} used users' neighbors on both platforms to establish profile similarities. 
Apart from the above mentioned content-dependent methods, several unsupervised models have utilized network structure to asses profile affinity \cite{heimann2018regal,derr2021deep,nassar2018low}.
 


\section{Discussion}
\label{sec:discussion}

Contemporary common belief assumes that using encrypted platforms provides sufficient privacy guarantees. 
Indeed, many individuals would feel secure using platforms that ensure neither the contents of their activity nor their interaction network leeks to other parties. 
Cross-domain privacy is often perceived as even safer, as users can intentionally vary or omit profile characteristics across platforms in order to enhance their privacy.

In this study, we demonstrate that the aforementioned privacy preserving means, adopted by both platform owners and users, are limited. 
Specifically, we show that temporal data, which is commonly collected by a variety of applications and is frequently shared without considering potential consequences, is highly sensitive.
Our proposed methodology enables identifying users across multiple anonymous domains, using merely the timing of their activities.
It establishes an encouraging performance that exceeds both naive temporal approaches and previously studied models, based on interaction networks. 

We postulate that the evident improvement over structure-based models arises from the fundamental difference between structural and temporal user characteristics. 
While the timing of a person's interactions depends only on the individual (at least explicitly), the structure is influenced not only by the activity of their immediate neighbours, but also by their k-hop neighbours. 
In problems involving multiple domains, with varied objectives and purposes, an individual's neighbourhood often changes. 
Indeed, consider our diverse networks in life: the people we trade with may differ from the people we socialize with for leisure, and these may differ from our professional contacts.
As a result, the surrounding structure of an individual might vary significantly across distinct domains. 

However, the timing of our activities presents a a more consistent fingerprint, even across different domains. 
While, as we have demonstrated, the actual timing of the different activities might not overlap, the intervals between these activities adhere to a unique individual pattern. 
This can be illustrated by the following example: an individual may use their free time between work sessions for different activities. This might include interacting on social platforms on day $1$, and trading crypto-currencies on day $2$. 
Though these two different activities do not overlap, the inter-event times between them might exhibit sufficient similarity, as they are likely to be influenced by the individual's regular work-leisure schedule characteristics.

We further assessed the model's robustness to noise and its impact on the individual temporal fingerprints. 
While adding various amounts of noise to individual transaction times affected the inter-event time distributions of true synchronized profiles, they remained rather similar even after noise injection.
Although this affected the predictive ability of the model, it still successfully linked $78\%$ of the matching profiles when examining at most $10$ profiles for each candidate, even with Gaussian noise with $\sigma ^2=1$ hour to each of the original transaction times.
The inherent robustness to noise sheds light on the limitations of applying state-of-the-art privacy preserving methodologies \cite{dwork2006calibrating,kairouz2014extremal} to temporal data.
Other strategies might incorporate state-of-the-art cryptographic solutions designed to obfuscate the user's identity, including methods like ring signatures \cite{van2013cryptonote} and zero-knowledge proofs \cite{sasson2014zerocash,movsowitz2023privacy}.

\subparagraph*{Limitations and future work}
Heavy tailed inter-event time distributions have been shown to characterize individual activity in a variety of other domains, ranging from individual mobility patterns, e-mail communications, instant messaging, web browsing and mobile phone calls, as presented in the works of  \cite{barabasi2005origin,gonzalez2008understanding, karsai2011small,oliveira2005darwin, zhou2008role}.
In this study we focus only on users' identity matching via their financial activity over the Ethereum blockchain. 
This decision stems from the practical advantages facilitated by the Ethereum blockchain in terms of performance evaluation.
Specifiaclly, the Ethereum blockchain enables using the same wallet for trading a variety of different crypto-coins, making it an ideal test-bed for assessing the effectiveness of identity matching techniques across the trading of these different coins. 
Nevertheless, establishing labelled datasets from other, non-financial, domains is crucial for verifying the generalization ability of such temporal techniques. 

Additional limitation is the computation complexity. 
Given an individual profile $u_{d_i}$ in domain $D_i$, the complexity of calculating their inter-event distribution in $O(m)$ where $m $ is the profile's activity volume. 
Calculating the Kolmogorov Smirnov test for a given profile pair requires sorting their inter-event times, and as such it takes $O(m_i \log m_i)$, where $m_i$ is the longest sequence of the two considered inter-event time sequences, without loss of generality.
Given $n$ domains, $k$ users in each the KS computation for the entire population will take $O(n^2 \times k^2 \times m_{\max}\log m_{\max})$.   
Possible solutions might entail employing complexity reduced versions of the Kolmogorov-Smirnov test, such as suggested by \cite{gonzalez1977efficient}. 
Improving the search algorithm beyond the naive bruit-force version we have employed could enhance computational efficiency as well. 

Finally, the inherent requirement for a \textit{sufficient} amount of individual data is another limitation of our study. 
In our analysis we have set the bar to a minimum of $20$ daily activities, and concentrated on daily inter-event distributions. 
As has been shown in Fig. 
in the appendix, the probability to identify multiple profiles pertaining to an individual increases with the volume of their activity.
Obtaining a broader comprehension of this identity matching mechanism heavily relies on further examining the dependence of the identification probability on individual activity volume and the length of time period on which the inter-event distribution is calculated.
Additionally, the choice of the examined similarity measure has a direct impact on the identity matching performance, and other methods, including those presented by \cite{scholz1987k} could be evaluated. 
The authors intend to pursue these directions in a future research.

\paragraph{Funding Acknowledgement}
Research was sponsored by the United States Air Force Research Laboratory and the Department of the Air Force Artificial Intelligence Accelerator and was accomplished under Cooperative Agreement Number FA8750- 19-2-1000. The views and conclusions contained in this document are those of the authors and should not be interpreted as representing the official policies, either expressed or implied, of the Department of the Air Force or the U.S. Government. The U.S. Government is authorized to reproduce and distribute reprints for Government purposes notwithstanding any copyright notation herein.


\appendix


\newpage

\bibliographystyle{ieeetr}

\bibliography{federated_anomalies}

\newpage
\section{Appendix}

%
%
%
%
%
%
%

\subsection{Transaction volume affect}\label{sec:trans_vol}
\begin{figure}[H]
		\centering
		{\includegraphics[height=6.1cm]{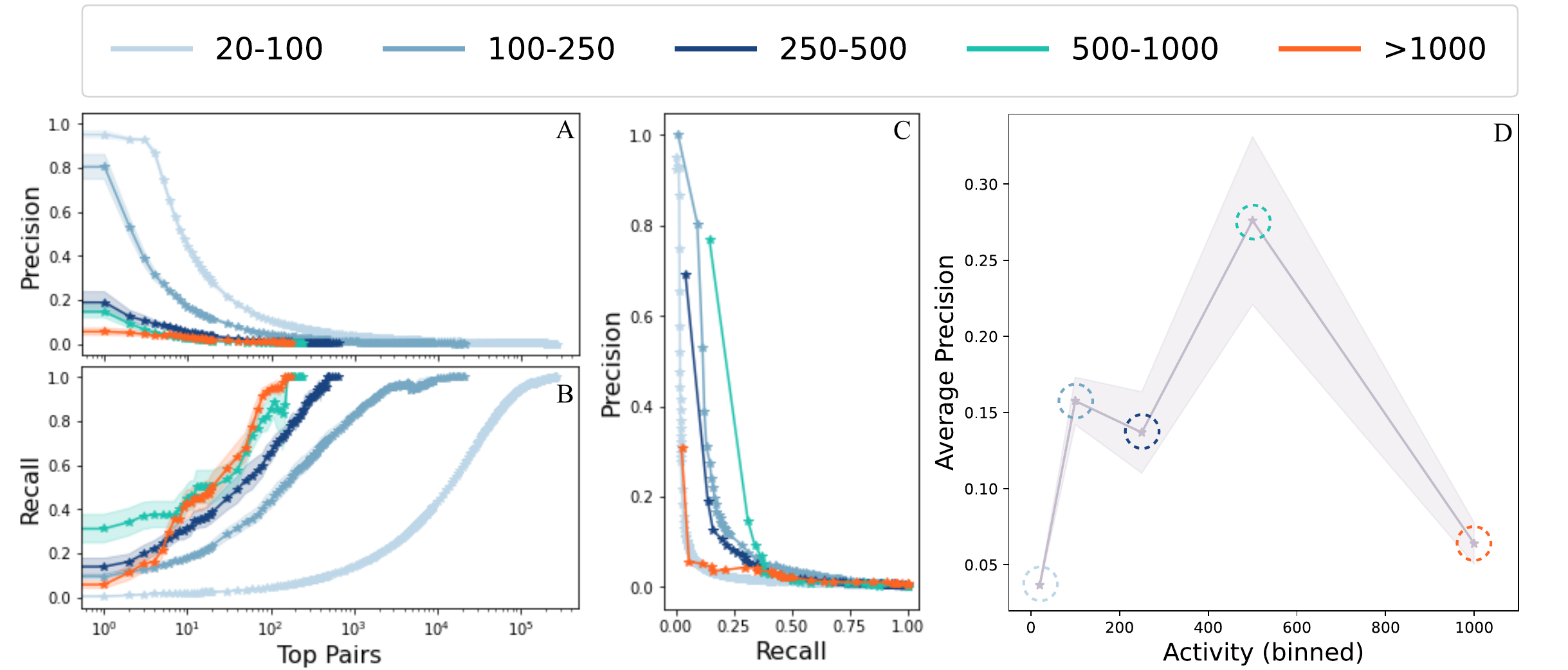} }
		\caption{Effect of profile transaction volume on identity matching performance. Panels A and B correspondingly present the precision and recall of the inter-event identity matching function of each of the activity categories. 
		Panel C depicts the precision-recall curves for each category. 
		Panel D presents the associated averaged precision (AP), where marker colors indicate the category type.  
 The results are averaged values over $14$ test days are presented by the solid lines, alongside with $\pm 1$ standard error in light background.
 This analysis indicates that precision decreases with activity volume while recall increases with it. The medium-volume activity category presents the highest AP. 
	}
		\label{fig:fig_si1_recision_recall_activity}
	\end{figure}

We wish to examine the affect of profiles' transaction volume on the ability to accurately associate them with the same individual. 
To this end, we divide the profiles into five categories according to their transactional volume:
\begin{equation}
\mathcal{C}^{i}_{\tau}(l,h)=\left\{ u_{d_i}\in D^i_{\tau} \biggr| l \leq |A_{\tau}^{u_{d_i}}| < h  \right\}
\end{equation}
where $l,\,h \in [20,100,250,500,1000]$.
We define the projection of $p^{ks}$ on each of these profile categories as follows:
\begin{equation}
p^{ks}_{\mathcal{C}}: \bigcup_{i,j \in [n]}\mathcal{C}^{i} \times \mathcal{C}^{j} \rightarrow [0,1]
\end{equation}
Fig. \ref{fig:fig_si1_recision_recall_activity} depicts the precision and recall of  $p^{ks}_{\mathcal{C}}$ for each activity category $\mathcal{C}$.
Panels A and B present the precision and recall on each activity category, manifesting the trade-off between precision and recall. 
Specifically, while precision decreases with the transactional volume, it is notable that the recall increases. 
Panel C presents the associated precision-recall curves. Panel D depicts the averaged precision (AP, the area under the precision-recall curve) as a function of the profiles' activity category:
\begin{equation}
AP = \int_{r=0}^{1} p(r) \, dr
\end{equation}
$r$ being the recall and $p(r)$ the precision at that point. 
Interestingly, the AP is not a monotonic function of the activity volume, maximized for profile pairs with medium transactional volume.  

Thus far, we have concentrated on the general identity matching performance, namely the ability of the presented algorithm to correctly identify pairs of profiles pertaining to the same individual across different domains. 
We would further wish to examine a more individual perspective. 
Specifically, we wish to estimate the probability to identify an individual on more that one domain, and its dependency on the individual's transactional volume.
We postulate that increased individual activity leads to higher detection rates. 
We define the set of all \textit{synchronized} correct matches as the profile pairs pertaining to same individual and also obtaining a KS distance lower that a pre-determined threshold:
\begin{equation}
    \mathcal{S}_{\tau}^{\rho} = \Bigl\{(u_{d_i},v_{d_j})\in \bigcup_{i,j\in [n]}D^i _{\tau}\times D^j_{\tau} \,\, \biggr| \,\, 
(KS_{\tau}(u_{d_i},v_{d_j})\leq \rho) \wedge  (u_{d_i},v_{d_j}) \in \mathcal{P}_{\tau} \Bigr\}
\end{equation}
We examine an individual's identification probability $P_{id}$ conditioned on their transactional volume:
\begin{equation}
P_{id}^{\rho,m} = Pr \left(u_{d_i} \in  \mathcal{S}_{\tau}^{\rho}\biggr| \,\,\,  |A^{u_d}_{\tau}| \approx m \,\, \wedge \,\, u_{d_i} \in \mathcal{P}_{\tau} \right)
\end{equation}

where $u_{d_i} \in  \mathcal{S}_{\tau}^{\rho}$ signifies $\exists\,v_{d_j} \in \bigcup_{j}D^j  \text{  s.t. } (u_{d_i},v_{d_j})\in \mathcal{S}_{\tau}^{\rho}$, and analogically for $u_{d_i} \in  \mathcal{P}_{\tau}$.

\begin{figure}[H]
		\centering
		{\includegraphics[height=7.4cm]{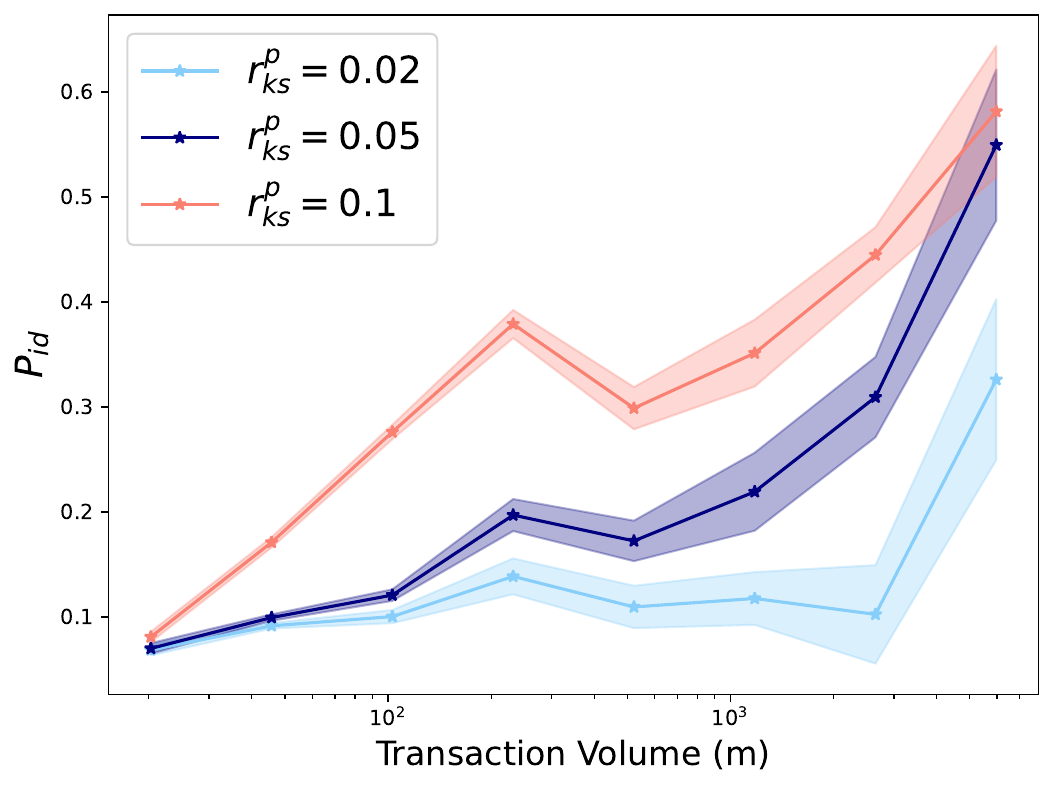} }
		\caption{Probability to correctly identify an individual conditioned on their transactional volume and the chosen synchronization threshold. 
The probability of correct identification increases with the transactional volume across all three examined synchronization thresholds  $\rho \in [0.02,0.05, 0.1]$, reaching over $50\%$ for $\rho=0.05$ and $m \approx 4000$. 
Averaged values over $14$ test days are presented by the solid lines, alongside with $\pm 1$ standard error in light background.
	}
		\label{fig:fig2_si_identification_prob_activity}
	\end{figure}

Fig. \ref{fig:fig2_si_identification_prob_activity} shows that the averaged identification probability increases monotonically with the transactional volume.
For instance, $P_{id}^{\rho,m}$ reaches $50\%$ for $\rho=0.05$ and $m \approx 4000$, indicating to the high probability of high-volume actors being identified by the suggested approach. 

\subsection{Stability}\label{sec:SI_stability}
We wish to estimate the stability of the inter-event synchronization.
First, we evaluate how the synchronization of two profiles which correspond to the same individual changes along time.
Concretely, we consider $ \mathcal{S}_{\tau}^{\rho}$ for a predefined day $\tau = t_0$ and $\rho=0.02$.
For a any given pair $(u_{d_i},v_{d_j})\in  \mathcal{S}_{t_0}^{\rho}$ and day $t \in \left[t_0,t_0+365\right]$ we consider $KS_{t}(u_{d_i},v_{d_j})$, the Kolmogorov-Smirnov distance of their corresponding inter-event time distributions at time $t$.
We evaluate the distribution of all computed Kolmogorov-Smirnov distances:
\begin{equation}
    \bigcup_{ \mathcal{S}_{t_0}^{\rho}}\bigcup_{t} KS_{t}(u_{d_i},v_{d_j})
\end{equation}

\begin{figure}[H]
		\centering
		{\includegraphics[height=4.4cm]{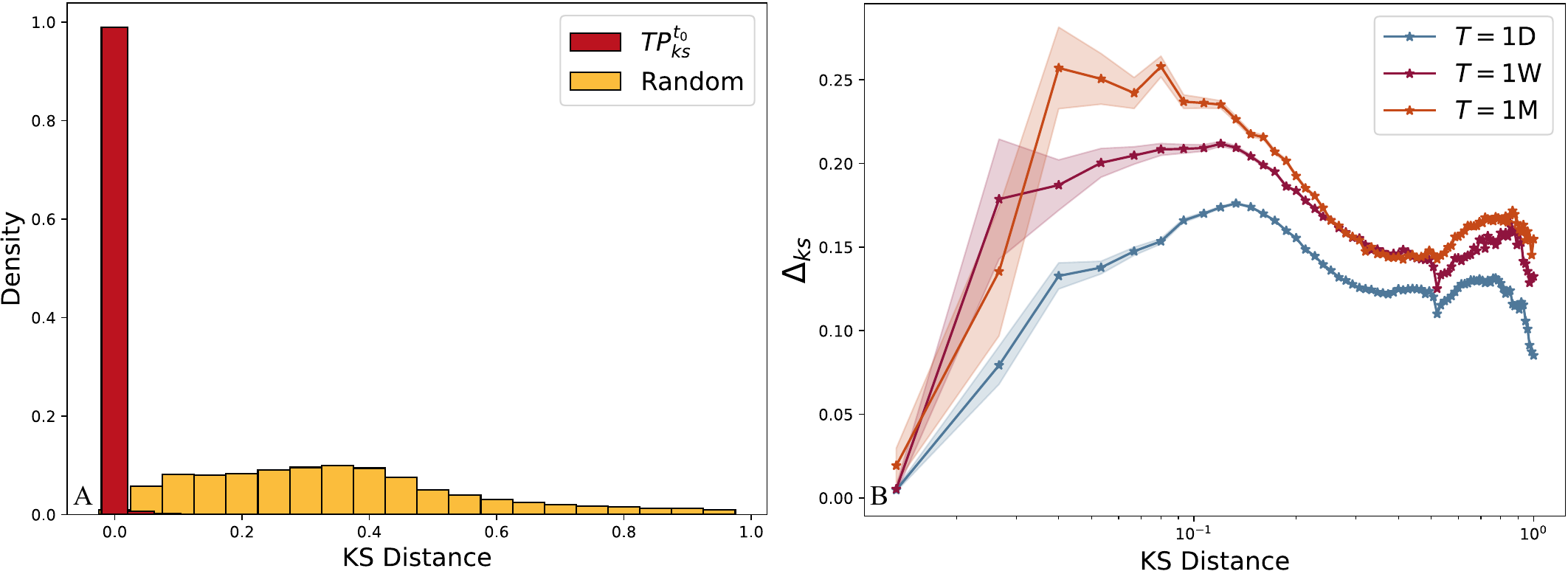} }
		\caption{Stability evaluation of inter-event distributions.
        Panel A presents the distribution of daily KS distances of synchronized profiles $TP^{t_0}_{ks}$ along an entire year (red bars).
It is compared to the KS distribution of a random, activity preserving, set (yellow bars), demonstrating that different profiles which correspond to the same individual manifest significantly greater synchronization stability, even throughout long periods of time. 
        Panel B depicts the dependence of similarity changes $\Delta_{ks}$ (average, with $\pm 1$ standard error in light background) on the initial similarity between different profiles, indicating that both highly similar profiles and highly distinct profiles are likely to be more stable along time.
	}
		\label{fig:fig1_si_stability}
	\end{figure}

Panel A in Fig. \ref{fig:fig1_si_stability} presents the distribution of daily KS distances established for profile pairs in $ \mathcal{S}_{t_0}^{\rho}$ compared to a random set\footnote{Randomly selected profiles preserving the transactional volume in the corresponding domain.}, calculated throughout an entire year (depicted by red bars and yellow bars respectively).
Encouragingly, the KS distances for $ \mathcal{S}_{t_0}^{\rho}$ present a highly skewed distribution, with significantly lower KS values throughout the examined period, as opposed to the higher and more varied KS values established by the random set. 
This indicates to the stability in the synchronization of different profiles corresponding to the same individual, even throughout long periods of time. 

Next, we analyse the overall stability of the distance between inter-event distributions and its dependence on the initial similarity between different profiles. 
Specifically, given a profiles pair $(u_{d_i},v_{d_j})\in D^i \times D^j$ and their corresponding inter-event time distributions over a 1-day period $[t_0,t_0+1]$, we wish to verify how their Kolmogorov-Smirnov distance changes along time. 
We define the difference of the KS distributions along an examined time-span $T$ as: 
    \begin{equation}
        \Delta_{ks}^{T} = \left|KS_{t_0+ T}(u_{d_i},v_{d_j})-KS_{t_0}(u_{d_i},v_{d_j})\right|
    \end{equation} 
Panel B in Fig. \ref{fig:fig1_si_stability} presents a non-monotonic dependence of $\Delta_{ks}^{T}$ on the initial $KS$ similarity for different periods $T$.
This indicates that both highly similar profiles and highly distinct profiles are likely to be more stable along time, reflecting on the overall stability of this similarity measure, as profile-pairs are unlikely to transition between these extremes, even throughout long periods of time.
	
\section{Temporal graph neural network}
The test case dataset includes all ERC20 token transactions on the Ethereum network recorded over 13 days, from May 6 to May 18, 2022. 
From these daily transaction networks we induce the corresponding similarity networks $G^{\tau}_{ks}$. 
Edges between nodes are defined by the daily similarity in their inter-event time distribution, as measured by the Kolmogorov–Smirnov (KS) test, where a lower KS score corresponds to higher similarity.
We frame this problem as a link prediction task, aiming to unveil potential connections within these similarity networks over time. 
Specifically, the task is identifying which profile pairs should be linked as corresponding to the same individual. 
This approach aligns with our objective to identify the implicit links between profiles, represented by edges in the similarity network, rather than to predict future transaction activity, as would be the case if we performed link prediction on the original transaction network.

For each profile, we calculate 16 node features, expanded to 34 after binning. Features include transaction counts, inter-event time statistics, and graph-based metrics like in-degree, out-degree, clustering coefficient, closeness, betweenness centrality, and PageRank. 
We use RobustScaler for feature normalization to mitigate the effects of outliers.
Node features are recalculated daily to reflect the evolution of transaction behavior, providing a consistent set of features for training, validation, and evaluation.

\subsection{Model Architecture}

We implemented a TGNN using the PyTorch Geometric Temporal library to analyze dynamic network data~\cite{rozemberczki2021pytorch}. The core of the architecture included:

\begin{enumerate}
    \item \textbf{GConvGRU Layer:} This layer combines the functionalities of Gated Recurrent Units (GRUs) with graph convolution operations to interpret temporal dependencies within the graph structure~\cite{seo2018structured, cho2014learning}. The update gate in the GConvGRU layer determines how much of the past information to retain:
    \begin{equation}
        z_t = \sigma(\mathbf{W}_z \cdot [h_{t-1}, x_t] + \mathbf{b}_z)
    \end{equation}
    where \( z_t \) is the update gate vector, \( \sigma \) is the sigmoid function, \( \mathbf{W}_z \) is the weight matrix for the update gate, \( h_{t-1} \) is the hidden state from the previous time step, \( x_t \) is the input feature vector at the current time step, and \( \mathbf{b}_z \) is the bias vector for the update gate.

    The reset gate controls the degree to which the past state influences the current state:
    \begin{equation}
        r_t = \sigma(\mathbf{W}_r \cdot [h_{t-1}, x_t] + \mathbf{b}_r)
    \end{equation}
    where \( r_t \) is the reset gate vector, \( \mathbf{W}_r \) is the weight matrix for the reset gate, and \( \mathbf{b}_r \) is the bias vector for the reset gate.

    The new memory content in the GConvGRU layer is calculated using the reset gate:
    \begin{equation}
        \tilde{h}_t = \text{tanh}(\mathbf{W} \cdot [r_t \odot h_{t-1}, x_t] + \mathbf{b})
    \end{equation}
    where \( \tilde{h}_t \) is the candidate hidden state, \( \text{tanh} \) is the Tanh activation function, \( \mathbf{W} \) is the weight matrix, \( \odot \) denotes element-wise multiplication, and \( \mathbf{b} \) is the bias vector.

    The final memory at the current time step is a combination of the past state and the new memory content:
    \begin{equation}
        h_t = z_t \odot h_{t-1} + (1 - z_t) \odot \tilde{h}_t
    \end{equation}
    where \( h_t \) is the hidden state at the current time step.

    \item \textbf{GATConv Layer:} This layer refines features by weighting the importance of neighboring nodes through learned attention mechanisms~\cite{velickovic2017graph}. The attention coefficients in the GATConv layer determine the importance of neighboring nodes:
    \begin{equation}
        \alpha_{ij} = \frac{\exp(\text{LeakyReLU}(\mathbf{a}^T [\mathbf{W} h_i \parallel \mathbf{W} h_j]))}{\sum_{k \in \mathcal{N}(i)} \exp(\text{LeakyReLU}(\mathbf{a}^T [\mathbf{W} h_i \parallel \mathbf{W} h_k]))}
    \end{equation}
    where \( \alpha_{ij} \) is the attention coefficient for edge \( (i,j) \), \( \mathbf{a} \) is the attention weight vector, \( \mathbf{W} \) is the weight matrix, \( h_i \) and \( h_j \) are the hidden states of nodes \( i \) and \( j \), \( \parallel \) denotes concatenation, and \( \mathcal{N}(i) \) is the set of neighbors of node \( i \).

    The node embeddings are computed as a weighted sum of the neighboring node features:
    \begin{equation}
        h_i' = \sigma \left( \sum_{j \in \mathcal{N}(i)} \alpha_{ij} \mathbf{W} h_j \right)
    \end{equation}
    where \( h_i' \) is the updated hidden state of node \( i \), and \( \sigma \) is the non-linear activation function.

    \item \textbf{Linear Embedding Layer:} This layer reduces node representations into a lower-dimensional space. Node features were transformed to a hidden dimension of 64, then projected down to an embedding dimension of 32:
    \begin{equation}
        \mathbf{e} = \mathbf{W}_e \cdot h + \mathbf{b}_e
    \end{equation}
    where \( \mathbf{e} \) is the embedding vector, \( \mathbf{W}_e \) is the embedding weight matrix, and \( \mathbf{b}_e \) is the embedding bias vector.

    \item \textbf{Edge Classifier:} This component takes the 32-dimensional embeddings for each node, concatenates them to form a 64-dimensional edge feature vector, and processes it through a linear classifier layer to predict the likelihood of an edge linking the nodes:
    \begin{equation}
        \text{logits} = \mathbf{W}_c \cdot [\mathbf{e}_s \parallel \mathbf{e}_t] + \mathbf{b}_c
    \end{equation}
    where \( \text{logits} \) are the raw scores for the edge existence prediction, \( \mathbf{W}_c \) is the classifier weight matrix, \( \mathbf{e}_s \) and \( \mathbf{e}_t \) are the source and target node embeddings, \( \mathbf{e}_s \parallel \mathbf{e}_t\) denotes the concatenation of source and target embeddings, and \(\mathbf{b}_c\) is the bias vector for the classifier.

\end{enumerate}

\begin{figure}
    \centering
    \includegraphics[width=1.06\linewidth, trim=0 50 0 50, clip]{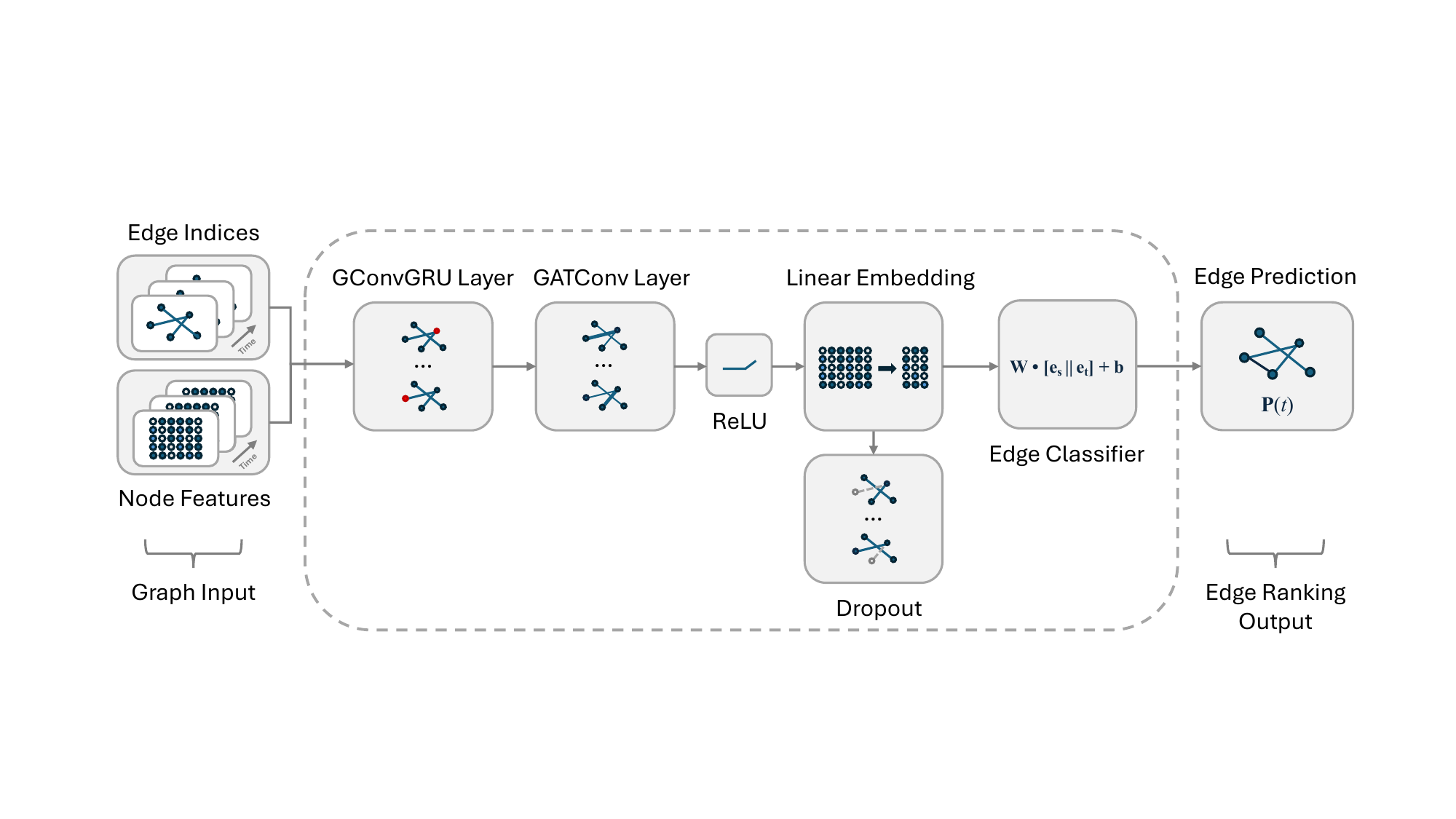}
    \caption{Architecture of the proposed Temporal Graph Neural Network (TGNN) model. The model processes graph input data, including edge indices and node features, through a series of layers. The GConvGRU layer captures temporal dependencies, followed by the GATConv layer which refines features using attention mechanisms. The output is then passed through a ReLU activation function and a Linear Embedding layer to reduce the dimensionality of node representations. Dropout is applied to prevent overfitting before edge classification. The final edge prediction step outputs the probability of edge existence, enabling edge ranking.}
    \label{fig:tgnn-arch}
\end{figure}

The model architecture is outlined in Figure~\ref{fig:tgnn-arch}.

\subsection{Training and Evaluation}

\textbf{Training:} We employed supervised learning to train the TGNN, using edge labels informed by KS distance metrics. Positive training edges were those with a KS score \(\leq 0.001\), and negative training edges were those with a KS score \(> 0.98\). A weighted binary cross-entropy loss function was used to counter class imbalance:
\begin{equation}
    \mathcal{L}_{\text{BCE}} = -\frac{1}{N} \sum_{i=1}^{N} \left[ w_i \cdot y_i \cdot \log(\sigma(\text{logits}_i)) + (1 - y_i) \cdot \log(1 - \sigma(\text{logits}_i)) \right]
\end{equation}
where \(\mathcal{L}_{\text{BCE}}\) is the binary cross-entropy loss, \(N\) is the number of samples, \(w_i\) is the weight for sample \(i\), \(y_i\) is the true label, \(\sigma\) is the sigmoid function, and \(\text{logits}_i\) are the raw scores from the classifier.

The Adam optimizer with an initial learning rate of 0.0001 and a ReduceLROnPlateau scheduler was used. Training was set to a maximum of 200 epochs with early stopping. Data snapshots were split 80:20 for training and validation, with the initial 80\% (days one to ten) allocated for training and the remaining 20\% (days eleven to thirteen) for validation, to ensure the model did not overfit to the training data.

\textbf{Evaluation:} We assessed the TGNN's capability to link Ethereum profiles against the baseline inter-event time distribution similarity. Consistent node features were used, and evaluation was performed across all 13 days of wallet activity. The TGNN predicted the likelihood of an edge existing for each pair within two sets independently. First, previously seen edges, i.e. the 146 positive training edges. Then, the next 1,000 most likely edges as ranked by KS score, corresponding to edges with a KS distance measure between 0.001 and 0.02616. The second set of edges had not previously been seen by the model.

\end{document}